\documentclass[onecolumn,useAMS,usenatbib]{mn2e}

\usepackage{color}
\usepackage{amsmath}
\usepackage{graphicx}
\usepackage{amssymb}
\usepackage{psfrag}

\newcommand{\beq}{\begin{equation}}
\newcommand{\eeq}{\end{equation}}
\newcommand{\beqy}{\begin{eqnarray}}
\newcommand{\eeqy}{\end{eqnarray}}

\newcommand{\Bav}{\bar{B}}

\newcommand{\brac}[1]{\left( {#1} \right)}

\newcommand{\nn}{\nonumber}

\newcommand{\be}{{\bf{e}}}
\newcommand{\br}{{\bf{r}}}
\newcommand{\bv}{{\bf{v}}}
\newcommand{\dvmu}{\frac{1}{4\pi}}
\newcommand{\bOm}{{\boldsymbol{\Omega}}}
\newcommand{\bB}{{\bf B}}
\newcommand{\pd}[2]{\frac{\partial{#1}}{\partial{#2}}}
\newcommand{\curl}{\nabla\times}
\renewcommand{\div}{\nabla\cdot}

\newcommand{\dP}{\delta P}

\newcommand{\drh}{\delta\rho}
\newcommand{\dB}{\delta\bB}
\newcommand{\boldf}{{\bf f}}

\newcommand{\rmd}{{\mathrm d}}
\newcommand{\rme}{{\mathrm e}}

\title[Instabilities in NSs with toroidal fields]{Instabilities in
  neutron stars with toroidal magnetic fields}
\author[S. K. Lander and D. I. Jones]
       {S. K. Lander\thanks{skl@soton.ac.uk} and D. I. Jones\\
School of Mathematics, University of Southampton, Southampton SO17 1BJ}
\begin{document}


\pagerange{\pageref{firstpage}--\pageref{lastpage}} \pubyear{0000}

\maketitle

\label{firstpage}

\begin{abstract}
We study $m=1$ oscillations and instabilities of magnetised neutron
stars, by numerical time-evolution of linear perturbations of the
system. The background stars are stationary equilibrium configurations
with purely toroidal magnetic fields. We find that an $m=1$
instability of toroidal magnetic fields, already known from local
analyses, may also be found in our relatively low-resolution global
study. We present quantitative results for the instability growth
rate and its suppression by rotation. The instability is
discussed as a possible trigger mechanism for Soft Gamma Repeater (SGR)
flares. Although our primary focus is evolutions of magnetised stars, we also
consider perturbations about unmagnetised background stars in order to
study $m=1$ inertial modes. We track these modes up to break-up
frequency $\Omega_K$, extending known slow-rotation results.
\end{abstract}


\section{Introduction}

Magnetic fields have both stabilising and destabilising influences on
stars. These effects are particularly important when the magnetic
energy $M$ is a relatively large fraction of the gravitational energy
$W$. One such class of star is magnetars (with $M/|W|\sim 10^{-6}$),
an especially highly magnetised type of neutron star. Magnetars have
surface field strengths of around $10^{14}-10^{15}$ G, whilst it is
not unreasonable to expect their interior fields to be an order of
magnitude stronger still, 
i.e. up to around $10^{16}$ gauss; such a value for the field seems to
emerge from modelling of magnetar flares \citep{stella_highB} and
cooling \citep{kaminker_highB}. Such strong fields are likely to
influence the secular evolution of magnetars and could drive violent
dynamical-timescale events --- in particular, the giant flares of
SGRs. Strong internal magnetic fields could also 
help explain the nature of glitches in Anomalous X-ray Pulsars (AXPs)
\citep{dib}.

This paper is primarily concerned with magnetic instabilities in
magnetars, but a magnetic field can also help to reduce other classes of
instability. In the first seconds of its life a
nonmagnetic proto-neutron star may suffer convective instabilities, but 
\citet{miralles} showed that a magnetic field can reduce the effect of
these, and remove them altogether for field strengths
$B\gtrsim 10^{16}$ G; hence they are unlikely to operate in young
magnetars. As well as these convective instabilities, a rotating
perfect-fluid star is susceptible to gravitational-radiation driven
instabilities even if it rotates very slowly. In real neutron stars
viscosity is likely to stabilise slowly-rotating stars against this
effect, but it may still operate at higher rotation rates
\citep{nils_rmode}. In addition to viscosity, there are various
arguments to suggest that these 
instabilities may also be reduced in the presence of magnetic fields
\citep{rezz_rmode,glam_ander,tor_mode}.

As well as the stabilising effects, however, magnetic fields also
induce instabilities. \citet{taylertor} found that a large class of
purely toroidal 
magnetic fields are susceptible to instabilities that develop near
the magnetic axis of the star. The worst instabilities in the linear
regime are those with $m=1$, but they exist for all $m$
\citep{goostayler}. These Tayler instabilities seem to be reduced by
the effect of rotation \citep{braithtor}, but it is not clear whether
typical stellar rotation rates are high enough to provide complete
stabilisation \citep{pitts_tayler}.

The flares from SGRs and glitches of AXPs indicate that magnetars do
not exist in an exact stable equilibrium. In a scenario put forward by
\citet{duncthom}, magnetic fields could be wound up by a dynamo
process, resulting in a large toroidal component. Should magnetar fields
indeed be dominantly toroidal, then they could be on the verge of
instability. If there are processes that can push the magnetic field
into an unstable configuration, then the Tayler instability could be a
candidate trigger for SGR giant flares.

In this paper we describe the first study of the $m=1$ Tayler
instability for toroidal fields from a global evolution of
perturbations. Because this instability is expected to set in around
the magnetic axis, previous work involved a local treatment of the
problem \citep{taylertor,braithtor,kitch}; here we confirm that such
toroidal-field instabilities can be seen from a global analysis too. For
$m=0$ instabilities much progress was made by \citet{kiuchi_evol}, who
performed non-linear evolutions of relativistic stars with toroidal
fields. They found that the instability resulted in convective
motions, which rearranged the field into a stable configuration.

Prior to our work on the Tayler instability, we study $m=1$ modes of
an unmagnetised star. This provides us with a test of our code, as we
are able to compare our mode frequencies with those of
\citet{yoshlee}; we also extend these results to rapidly-rotating
configurations, up to around 95\% of Keplerian velocity. We then look
at $m=1$ perturbations of a magnetised star, finding the Tayler
instability described above. We quantify the growth of this
instability and its dependence on field strength, as well as the
degree to which it is suppressed by rotation. Finally we discuss the
possibility that this effect may provide the trigger for the giant
flares of SGRs.

\section{Governing equations and numerics}

In \citet{tor_mode} we investigated $m\geq 2$ oscillation modes of
neutron stars with toroidal background fields. Since we are concerned
with $m=1$ instabilities here, the only major changes in our set-up
are related to the azimuthal index. For this reason, we shall only
briefly review the perturbation equations and numerics, describing any
differences from the earlier paper.

We model a neutron star as a self-gravitating, rotating, magnetised
polytropic fluid with perfect conductivity. We wish to study linear
instabilities, for which the governing equations of the
system become a set of stationary background equations and a set of
equations describing the time evolution of the perturbations. The
background configuration has a purely toroidal magnetic field
$\bB_0=B_\phi\be_\phi$ and is rigidly rotating:
\beq \label{backeuler}
0 = -\nabla P_0 - \rho_0\nabla\Phi_0
    - \rho_0\bOm\times(\bOm\times\br) + \dvmu(\curl\bB_0)\times\bB_0,
\eeq
\beq
\nabla^2\Phi_0 = 4\pi G\rho_0,
\eeq
\beq
P_0=k\rho_0^\gamma.
\eeq
It can be shown that in axisymmetry a general toroidal field has the
form $B_\phi=\lambda\rho_0 r\sin\theta$, where $\lambda$ is a constant
governing the strength of the field; see \citet{landerjones} for
details. This form of field has been known since at least the work of
\citet{rox_tor}, who found that stellar magnetic fields had to be
dominated by this toroidal component to allow for analytically
tractable equilibria with meridional circulation. These background
equations are solved through an iterative 
procedure to find stationary equilibrium configurations; this is done
using the nonlinear code of \citet{landerjones} and allows for
distortions of the star due to rotational and magnetic effects.

For the perturbation equations, we work in the rotating frame of the
background and write our equations in terms of the perturbed density
$\drh$, the mass flux $\boldf=\rho_0\bv$ and a magnetic variable
$\bbeta=\rho_0\dB$. We additionally make the Cowling approximation ---
neglecting the perturbed gravitational force --- to avoid the
computational expense of solving the perturbed Poisson equation. Our
perturbations are then governed by seven equations:
\beq \label{euler_magmode}
\rho_0\pd{\boldf}{t}
  = -\gamma P_0\nabla\drh - 2\bOm\times\boldf
    + \brac{\frac{(2-\gamma)\gamma P_0}{\rho_0}\nabla\rho_0
                            - \dvmu(\curl\bB_0)\times\bB_0}\drh
    + \dvmu(\curl\bB_0)\times\bbeta + \dvmu(\curl\bbeta)\times\bB_0
    - \frac{1}{4\pi\rho_0}(\nabla\rho_0\times\bbeta)\times\bB_0,
\eeq
\beq \label{conti_magmode}
\pd{\drh}{t}=-\div\boldf,
\eeq
\beq \label{induc_magmode}
\pd{\bbeta}{t} = \curl(\boldf\times\bB_0)
                 -\frac{\nabla\rho_0}{\rho_0}\times(\boldf\times\bB_0).
\eeq
We set $\gamma=2$ as an approximation to neutron star matter. As in
\citet{tor_mode}, we perform an azimuthal decomposition of the
perturbation equations, allowing us to separate $m=1$ perturbations
from the full set, and reducing our problem from a 3D to a 2D one.

\subsection{Boundary conditions}

We simplify our surface treatment by replacing the radial coordinate
$r$ with one fitted to isopycnic surfaces, $x=x(r,\theta)$; even if
the stellar surface is nonspherical, it will still be defined by one
value $x\equiv R$. With the 
background density being a function of $x$ alone, we have
$\rho_0(x\!=\!R)=0$ and hence
\beq
\boldf(x\!=\!R)=\bbeta(x\!=\!R)={\bf 0}\ ,\ \delta P(x\!=\!R) = 0. 
\eeq
At the centre of the star we enforce a zero-displacement condition
(this condition is valid for all perturbations except $m=0$ ones):
\beq
\dP(x\!=\!0)=0\ ,\ \boldf(x\!=\!0)=\bbeta(x\!=\!0)={\bf 0}.
\eeq
For rotating stars with purely toroidal fields, the perturbation
variables have equatorial symmetry. Enforcing this means the numerical
domain can be reduced to just one 2D quadrant of a disc
\citep{tor_mode}.

The behaviour of perturbations at the pole may be deduced by using their
standard decompositions into axial and polar parts. A general vector
perturbation (the velocity is shown here) may be decomposed as
\beq \label{bv_decomp}
\bv = U(r)Y_{lm}\be_r + V(r)\nabla Y_{lm}
      + W(r)\be_r\times\nabla Y_{lm},
\eeq
whilst a scalar perturbation (in this case, the density) will have the
form
\beq \label{drh_decomp}
\drh = T(r)Y_{lm}.
\eeq
Although we do not decompose in $\theta$ in the code, we
will find it convenient to rewrite the spherical harmonics using
$Y_{lm}(\theta,\phi)\sim P_{lm}(\cos\theta)\rme^{im\phi}$ (the constants are
unimportant; they may be regarded as absorbed into the radial
function). The boundary conditions at the pole $\theta=0$ are then
given by the behaviour of the relevant functions of $P_{lm}$ there.
Using recurrence relations (see for example \cite{arfken}), one may
show that a Legendre function  
$P_{lm}$ contains a $\sin^m\theta$ term and that its
$\theta$-derivative $\rmd P_{lm}/\rmd\theta$ contains a
$\sin^{m+1}\theta$ term and a $\sin^{m-1}\theta$ term.

By \eqref{drh_decomp}, it is clear that scalar perturbations have
$\theta$-dependence given simply by $P_{lm}$; since we are concerned
with $m\neq 0$ perturbations our BC at the pole is that a scalar
perturbation must vanish there.
For vector perturbations, we first re-express \eqref{bv_decomp} in
terms of spherical polar components:
\beqy
v_r      &=& U(r)Y_{lm}, \\
v_\theta &=& V(r)\nabla Y_{lm}\cdot\be_\theta
             + W(r)(\be_r\times\nabla Y_{lm})\cdot\be_\theta \nn\\
         &=& \frac{\rme^{im\phi}}{r}
                \brac{V(r)P_{lm,\theta}-\frac{imW(r)}{\sin\theta}P_{lm}},\\
v_\phi   &=& V(r)\nabla Y_{lm}\cdot\be_\phi
             + W(r)(\be_r\times\nabla Y_{lm})\cdot\be_\phi \nn\\
         &=& \frac{\rme^{im\phi}}{r}
                \brac{V(r)P_{lm,\theta}+\frac{imW(r)}{\sin\theta}P_{lm}}.
\eeqy
From these, it is clear that $v_r=0$ at the pole for all $m\neq
0$. $v_\theta$ and $v_\phi$ may be expressed as powers of
$\sin\theta$ as described earlier; the lowest power in each case is
$\sin^{m-1}\theta$. We deduce that $v_\theta=v_\phi=0$ at the pole for
$m\geq 2$, whilst for $m=1$ the boundary condition only requires them
to be finite and continuous; in this case the boundary condition is
that the $\theta$-derivatives should vanish at the pole.

In summary, then, the boundary conditions at the pole for $\drh,\bv$
and $\bbeta$ are:
\beqy
\drh=v_r=\beta_r \!  &=&\!\! 0 \ \ \ \forall m\neq 0\ ;\\
v_\theta=v_\phi=\beta_\theta=\beta_\phi\!
                   &=&\!\! 0 \ \ \ (m\geq 2)\ ;\\
v_{\theta,\theta}=v_{\phi,\theta}
 =\beta_{\theta,\theta}=\beta_{\phi,\theta}\!
                   &=&\!\! 0 \ \ \ (m=1).
\eeqy      

\subsection{Initial data}

Because we do not decompose in $\theta$, arbitrary initial data with a
specified azimuthal index $m$ will excite modes containing $l\geq m$
harmonics. In principle there will be an infinite number of these, 
but on a grid with finite angular resolution only the lowest
are seen. Modes may then be identified by comparison with
slow-rotation results and from analysis of their eigenfunctions. We
choose different starting perturbations depending on whether we wish to
investigate axial/axial-led or polar/polar-led modes (using the
terminology of \citet{lock_fried}). All results for instabilities
presented in this 
paper have used axial initial data, but we find that similar results
may be obtained by using an initial perturbation which is polar.

\subsection{Numerics}

Having decomposed in $\phi$, and by enforcing boundary conditions at
the equator, surface, centre and pole, we need only study
perturbations on one quadrant of a (2D) disc. As described in
\citet{tor_mode}, this is done by time-evolving the perturbation
equations numerically, using a McCormack predictor-corrector
algorithm. We need to use artificial viscosity to remove
high-frequency numerical instabilities, but we take care to include
this at the minimum possible level, to avoid damping the physical
Tayler instability. For the same reason, we have not employed
artificial resistivity for the evolutions in this paper. To enforce
the solenoidal constraint $\div\dB=0$ we use a mixed
hyperbolic-parabolic divergence cleaning scheme, as described in
\citet{dedner} and \citet{pricemona}.

We nondimensionalise by dividing by a suitable combination of
gravitational constant $G$, equatorial radius $R$ and maximum
density $\rho_m$. Where we have used nondimensional variables these
are denoted with a hat; for example, the dimensionless rotation rate
(in radians per code time)
is written $\hat\Omega$, with
$\hat\Omega\equiv\Omega/\sqrt{G\rho_m}$. We redimensionalise to a
neutron star whose mass $\mathcal{M}_0$ and radius $R_0$ would take
the canonical values of 
$\mathcal{M}_0=1.4\mathcal{M}_\odot$ (where $\mathcal{M}_\odot$ is
solar mass) and $R_0=10$ km, 
if the star were nonrotating and unmagnetised (i.e. spherical and in
hydrostatic equilibrium). An approximate formula to convert
dimensionless frequencies ($\hat\Omega$ or $\hat\sigma$) to physical
ones is $\Omega[\mathrm{Hz}]\approx 1900\hat\Omega$.

\section{\MakeLowercase{$m=1$} modes in an unmagnetised star}
\label{m1modes_nomag}

Before looking at $m=1$ oscillations of magnetised stars, we first
wish to check our code reproduces known results for nonmagnetic
modes. Some of these may be of gravitational-wave interest; whilst
dipolar ($l=m=1$) modes do not radiate, higher-$l$ modes can.
\citet{yoshlee} included results for $m=1$ 
oscillations in their study of inertial modes of slowly rotating
stars. We therefore compare our results for slowly rotating stars with
their values, bearing in mind that we make the Cowling approximation,
which Yoshida and Lee do not. This could be expected to cause fairly
large errors in some cases, since the Cowling approximation is poorer
for low $m$. Notwithstanding these differences of approach, we find
convincing agreement with their work; see table \ref{YLcompare}. In
one case we would expect good agreement -- the $r$-mode, which is
purely axial in the slow-rotation limit. This mode frequency should
not be greatly affected by the Cowling approximation, and in this case
our result is only $0.6\%$ different from that of Yoshida and Lee.

\begin{table}
\begin{center}
\caption{\label{YLcompare}
         Comparison between Yoshida-Lee results and those from our
         time-evolution code run for a dimensionless rotation rate of
         $\hat{\Omega}=0.119$ ($\approx$17\% of Keplerian velocity
         $\Omega_K$). All mode frequencies are made 
         dimensionless through division by $\Omega$ and calculated in
         the rotating frame of the star. As in \citet{tor_mode}, our
         mode labelling is consistent with 
         that of \citet{lock_fried}. In Yoshida and
         Lee's results, corotating modes are shown with a negative mode
         frequency, whilst we are only 
         able to find the magnitude. Finally, we were unable to identify
         the ${}^3i_1$ mode, which we believe is due to its proximity in
         frequency space to the strong $r$-mode peak.}
\begin{tabular}{cccc}
mode &  frequency    &   frequency      & discrepancy \\
     & (Yoshida-Lee) & (time evolution) &             \\
\hline
${}^1r$   & 1.000   & 1.006 & 0.6\% \\
${}^2i_1$ & -0.4014 & 0.388 & 3.3\% \\
${}^2i_2$ & 1.413   & 1.418 & 0.4\% \\
${}^3i_1$ & -1.032  &   -   &  -    \\
${}^3i_2$ & 0.6906  & 0.684 & 1.0\% \\
${}^3i_3$ & 1.614   & 1.611 & 0.2\% \\
${}^4i_1$ & -1.312  & 1.241 & 5.4\% \\
${}^4i_2$ & -0.1788 & 0.171 & 4.2\% \\
${}^4i_3$ & 1.052   & 1.021 & 2.9\% \\
${}^4i_4$ & 1.726   & 1.738 & 0.7\%
\end{tabular}\\
\end{center}
\end{table}

One oddity of the $m=1$ spectrum is that there is no $f$-mode; a
dipolar mode with no radial node displaces the centre of mass of the
star.  However, if one makes the Cowling
approximation then an $f$-mode
\emph{does} appear in the frequency spectrum, in its usual place
between the (pressure) $p$-modes and the (gravity) $g$-modes. This
spurious mode shifts to become the lowest-order $g$-mode in the full
non-Cowling problem \citep{dipfmode}. 

\begin{figure}
\begin{center}
\begin{minipage}[c]{0.6\linewidth}
\psfrag{sigma}{$\hat\sigma$}
\psfrag{Om}{$\hat\Omega$}
\psfrag{1r}{${}^1r$}
\psfrag{3i2}{${}^3i_2$}
\psfrag{3i3}{${}^3i_3$}
\includegraphics[width=\linewidth]{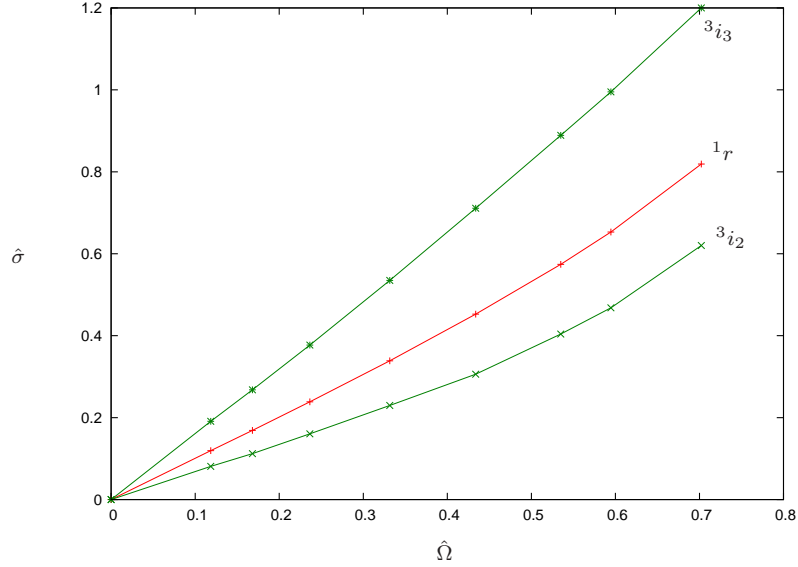}
\end{minipage}
\caption{\label{dip_ax}
         Axial-led $m=1$ inertial modes. The ${}^3i_1$ mode is
         missing; it seems to be obscured in the spectrum by the
         nearby $r$-mode, which has a very strong peak. Note that
         $\hat\Omega\equiv\Omega/\sqrt{G\rho_m}$ and
         $\hat\sigma\equiv\sigma/\sqrt{G\rho_m}$. We have tracked
         the modes up to rotation rates very close to the star's
         Keplerian velocity: $\hat{\Omega}_K\approx 0.72$ in
         these dimensionless units.}
\end{center}
\end{figure}

\begin{figure}
\begin{center}
\begin{minipage}[c]{0.6\linewidth}
\psfrag{sigma}{$\hat\sigma$}
\psfrag{Om}{$\hat\Omega$}
\psfrag{f}{$f$}
\psfrag{2i1}{${}^2i_1$}
\psfrag{2i2}{${}^2i_2$}
\psfrag{4i1}{${}^4i_1$}
\psfrag{4i2}{${}^4i_2$}
\psfrag{4i3}{${}^4i_3$}
\psfrag{4i4}{${}^4i_4$}
\includegraphics[width=\linewidth]{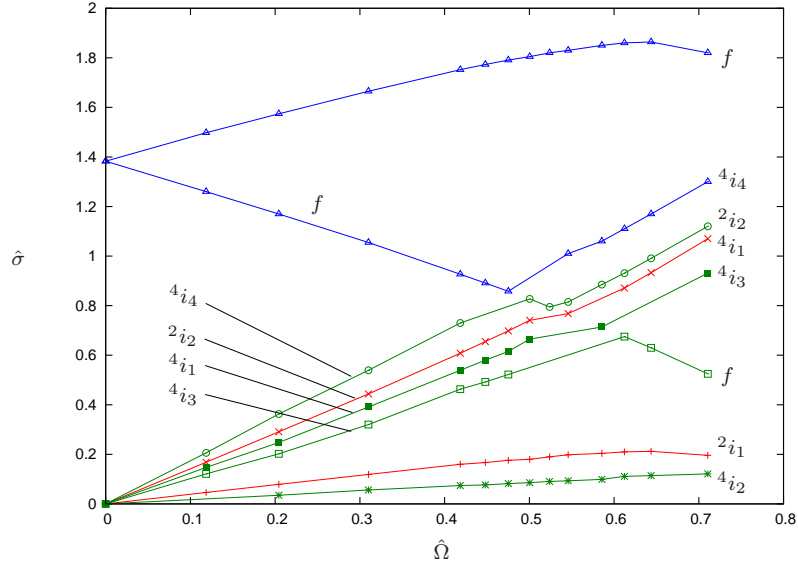}
\end{minipage}
\caption{\label{dip_pol}
         Polar-led $m=1$ inertial modes and the spurious $f$-mode,
         which has zero frequency in the full problem but appears as
         an oscillation mode of the Cowling-approximation system of
         equations. Four of the inertial modes have avoided crossings
         with the corotating branch of the $f$-mode, where their
         character changes; note the difference in labelling of these
         modes before and after the avoided crossings.}
\end{center}
\end{figure}

In addition to finding nine of the ten $m=1$ inertial modes described
by Yoshida and Lee, we also see the spurious $f$-mode described
above. Since our background configuration is generated in a nonlinear
manner, we are able to track the inertial modes up to break-up
velocity, where the results of Yoshida and Lee are no longer valid. We
also see avoided crossings between four of the polar inertial modes
and the corotating branch of the $f$-mode. These results are shown in
figures \ref{dip_ax} and \ref{dip_pol}.

\section{Instabilities in purely toroidal fields}

In the previous section we established that our time evolution code
ran stably for unmagnetised backgrounds with $m=1$, reproducing known
results for inertial modes as well as finding the spurious dipolar
$f$-mode that is an artefact of making the Cowling approximation. We
have, therefore, some confidence in the reliability of our $m=1$
evolutions of magnetised stars.

Before moving on to the results of our $m=1$ evolutions, let us review
what we can expect from previous studies. The stability analysis of
\citet{taylertor} established that a large class of toroidal field
configurations suffer localised instabilities; earlier calculations
than Tayler's had involved analysis of global integral quantities and
hence did not find evidence of the unstable nature of toroidal fields
(see, for example, \citet{rox_durn}). Tayler showed 
that instabilities tend to occur
close to the symmetry axis of the star, appearing over short
timescales (of the order of the Alfv\'en crossing time). Whilst $m=1$
perturbations appear to be the most unstable in the linear regime,
instabilities exist for all $m$ \citep{goostayler}. The instability is
reduced, but not necessarily eliminated, by rotation
\citep{pitts_tayler,braithtor,kiuchi_evol,kitch}. 

These studies into toroidal-field instabilities contrast with the work
reported in \citet{tor_mode}, where we were able to time-evolve
perturbations on a purely toroidal background field over long times
without seeing evidence of unstable oscillations; however, these
evolutions were for azimuthal indices $m\geq 2$ rather than the most
unstable $m=1$ perturbations. In addition, we included artificial
resistivity to remove numerical instabilities, but it may have also
damped out the genuine instability inherent in purely toroidal fields.

\begin{figure}
\begin{minipage}[c]{0.48\linewidth}
\psfrag{emag}{$\delta\hat{M}$}
\psfrag{t}{$\hat{t}$}
\psfrag{low}{low}
\psfrag{medium}{medium}
\psfrag{high}{high}
\includegraphics[width=\linewidth]{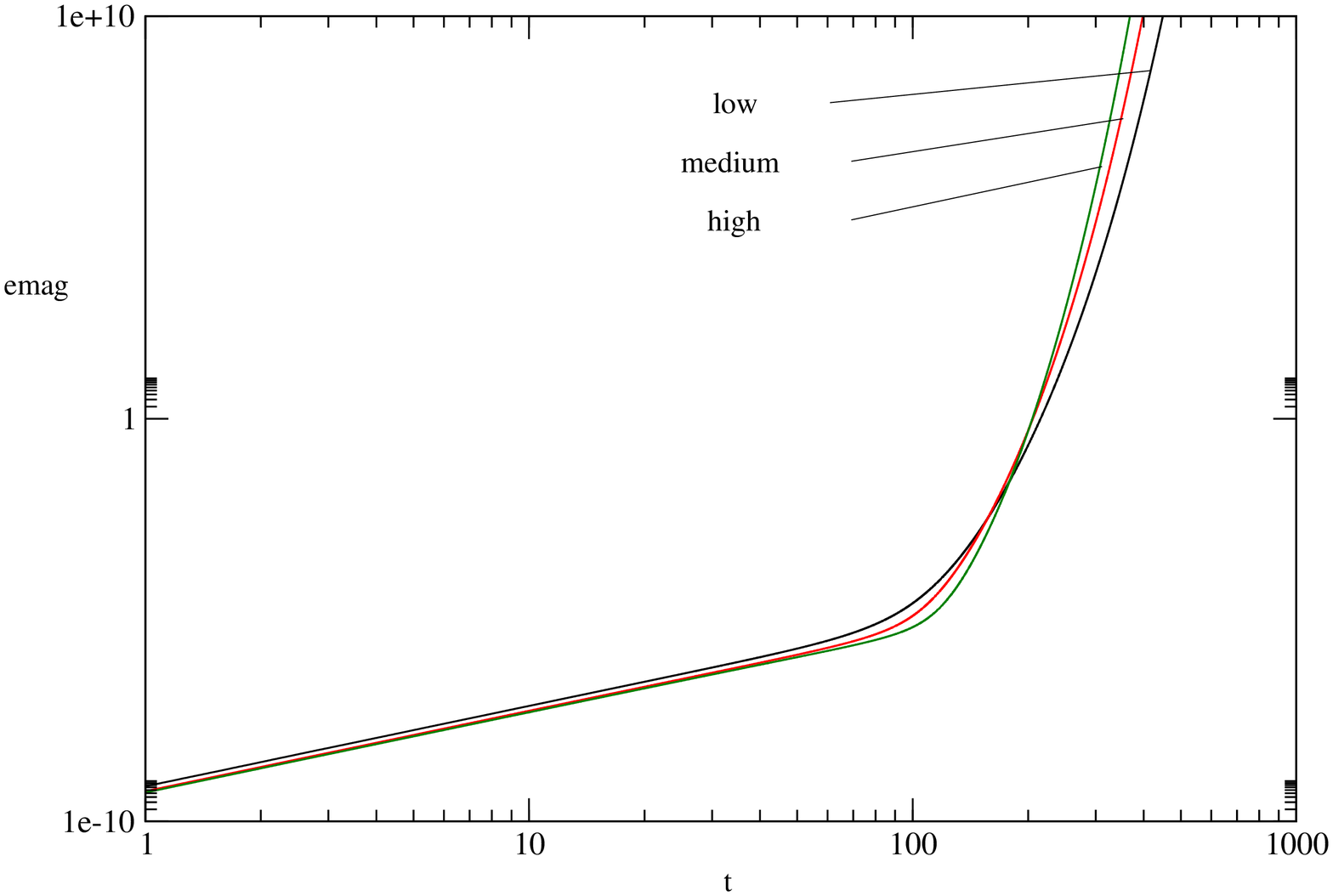}
\caption{\label{instab_norot}
         The Tayler instability for toroidal magnetic fields in a
         nonrotating star. We plot the magnetic energy
         $\delta\hat{M}$ against time $\hat{t}$, both in
         dimensionless form, for three different grid resolutions. We
         see that the onset time for the instability is independent of
         resolution (appearing at around the expected value of
         $\hat{\tau}_A\approx 77$), and its growth rate converges,
         suggesting that it may indeed be a physical instability. The
         results are for a star with average field strength
         $\Bav=3.0\times 10^{16}$ G.}
\end{minipage}
\hfill
\begin{minipage}[c]{0.48\linewidth}
\psfrag{emag}{$\delta\hat{M}$}
\psfrag{t}{$\hat{t}$}
\psfrag{low}{$\Bav=1.5\times 10^{16}$ G}
\psfrag{medium}{$\Bav=3.0\times 10^{16}$ G}
\psfrag{high}{$\Bav=6.0\times 10^{16}$ G}
\includegraphics[width=\linewidth]{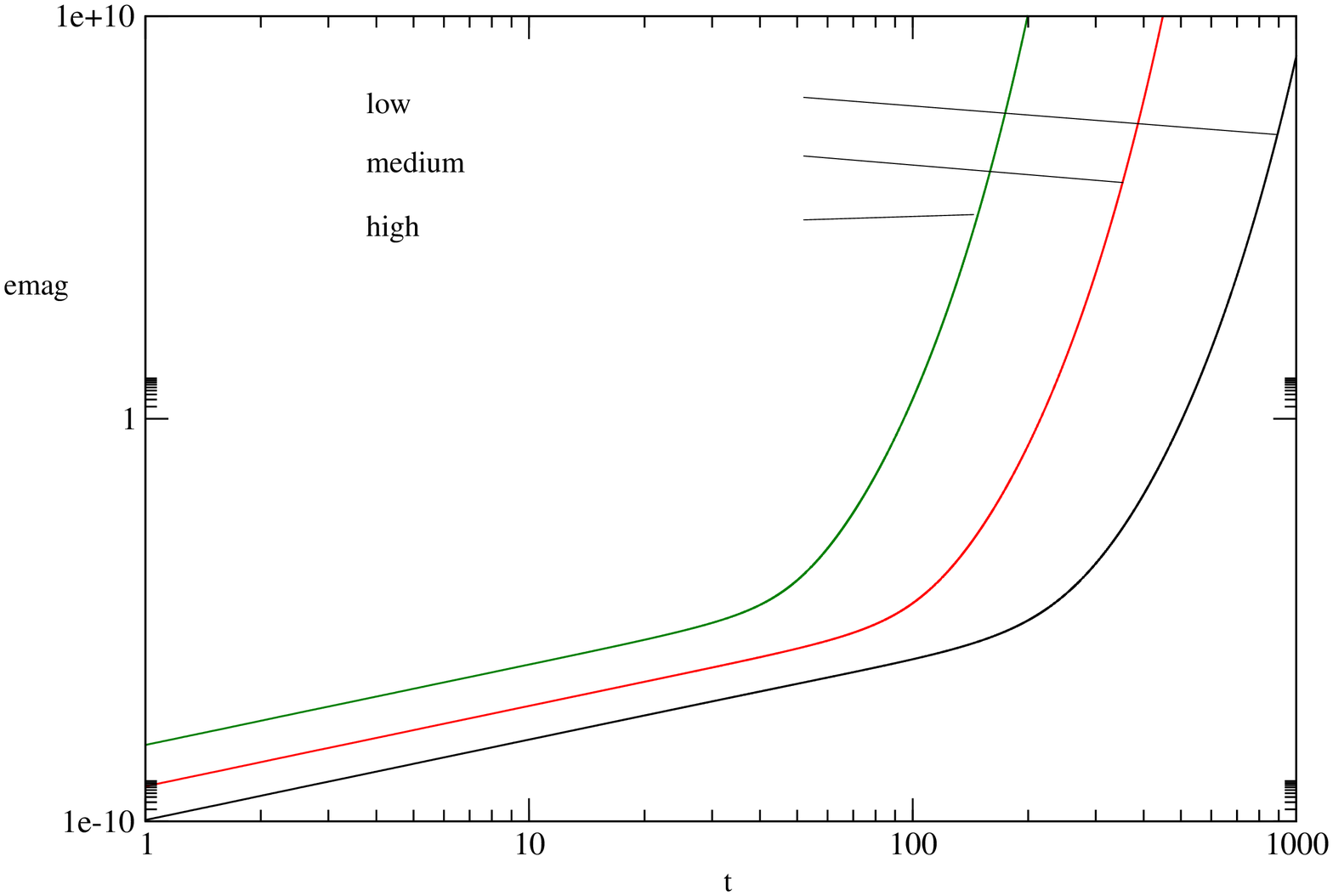}
\caption{\label{instab_varyB}
         Showing that the Tayler instability sets in after one
         Alfv\'en crossing time. We plot the magnetic energy against
         time, as before, and find that the onset of instability
         happens sooner for higher field strengths; in particular, the
         observed onset time in each case seems to be close to the
         Alfv\'en crossing time: $\hat{\tau}_A\approx 154,77,39$ for
         $\Bav=1.5,3.0,6.0\times 10^{16}$ G respectively.}
\end{minipage}
\end{figure}

\begin{figure}
\begin{center}
\begin{minipage}[c]{0.6\linewidth}
\psfrag{emag}{$\delta\hat{M}$}
\psfrag{t}{$\hat{t}$}
\psfrag{om0}{$\hat{\Omega}=0.0$}
\psfrag{om122}{$\hat{\Omega}=0.122\approx 0.17\hat{\Omega}_K$}
\psfrag{om237}{$\hat{\Omega}=0.237\approx 0.33\hat{\Omega}_K$}
\includegraphics[width=\linewidth]{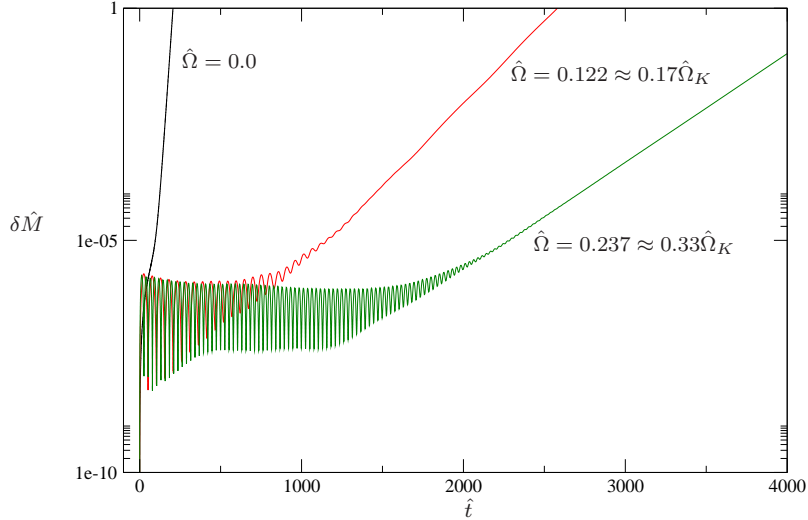}
\end{minipage}
\caption{\label{instab_rot}
         The stabilising effect of rotation on purely toroidal
         magnetic fields. The magnetic energy is plotted against time
         for three different rotation rates. We see that increasing
         the rotation rate decreases the growth rate of the
         instability; i.e. the gradient of $\delta M$ is reduced in
         the regime where the instability dominates. As for the
         previous plot, each configuration has a field strength of
         $\Bav=3.0\times 10^{16}$ G.}
\end{center}
\end{figure}

We first consider $m=1$ perturbations of nonrotating stars with
toroidal magnetic fields. Monitoring the perturbed magnetic energy
$\delta M=\int (\delta B)^2/8\pi \ \rmd V$ over time, we see exponential
growth of the perturbation, indicating instability. To check that this
is representative 
of the physical system and not a numerical instability, we perform a
convergence test; see figure \ref{instab_norot}. We compare evolutions
for three different grid resolutions: low (16 radial points $\times$
15 angular ones),
medium ($32\times 30$) and high ($64\times 60$). In all cases the
instability seems to set in at the same point; whilst by comparing the
three gradients we find that the growth rate converges with resolution
at approximately second order (the intended accuracy of the code).

\citet{taylertor} suggests that the toroidal-field instability
uncovered in his work should appear after approximately one Alfv\'en
crossing time, i.e. after
\beq
\tau_A \approx \frac{2R}{<\!c_A\!>}
          =     2R \sqrt{\frac{4\pi<\!\rho\!>}{\Bav^2}},
\eeq
where $R$ is the stellar radius, $c_A$ the Alfv\'en speed and angle
brackets denote volume averages. Evaluating this
in dimensionless form for a star with average field strength $\Bav=3.0\times
10^{16}$ G gives $\hat{\tau}_A\approx 77$; this is consistent with
the results shown in figure \ref{instab_norot}, where $\delta M$ is
seen to begin growing rapidly at $\hat{t}\approx 80-100$. To check
that this is not a coincidence, we plot the results for three
different field strengths in figure \ref{instab_varyB}. As expected,
in each case the instability appears to set in after one Alfv\'en
crossing time. 

Further evidence that we are seeing the Tayler instability is the
behaviour of our $m=1$ toroidal-field evolutions in the presence of
rotation. This is expected to reduce the effect of the Tayler
instability \citep{pitts_tayler}, which is what we find. In figure
\ref{instab_rot} we 
compare the behaviour of $\delta M$ in rotating and nonrotating
evolutions. We see that in the presence of rotation the system
develops oscillations and the instability does not manifest itself
until considerably later times; from this we deduce that,
qualitatively, the growth of the instability has been slowed by
rotation. We will return to study the instability quantitatively later
in this section.

\begin{figure}
\begin{minipage}[c]{0.48\linewidth}
\begin{center}
\includegraphics[width=0.7\linewidth]{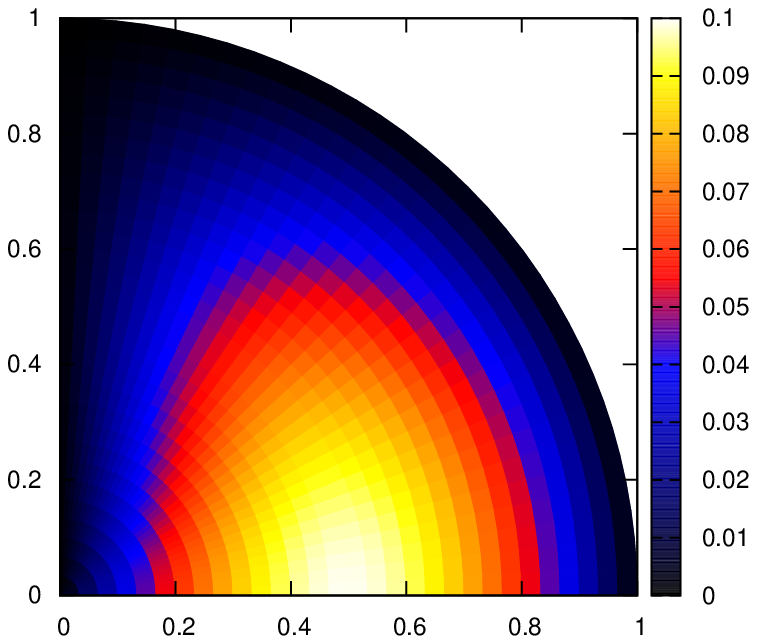}
\caption{\label{B_back}
         The toroidal field configuration of the background star. Note
         that the field vanishes on the pole and at the surface,
         reaching a maximum at a dimensionless radius $r/R \sim 0.5$
         from the centre.}
\end{center}
\end{minipage}
\hfill
\begin{minipage}[c]{0.48\linewidth}
\begin{center}
\includegraphics[width=0.7\linewidth]{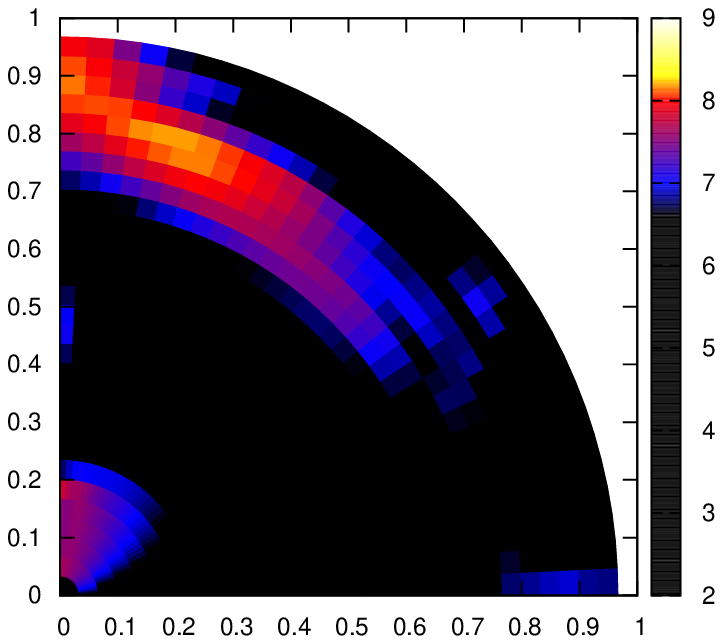}
\caption{\label{B_pert}
         The magnitude of the perturbed magnetic field after
         the onset of instability. The most unstable perturbations are
         visible around the magnetic axis and in a region
         close to the stellar surface, where the background field is
         weak. The growth of the instability is shown by dividing the
         perturbed field through by an early-time 
         perturbed field configuration.}
\end{center}
\end{minipage}
\end{figure}

One of the predictions made by \citet{taylertor} is that an $m=1$
toroidal-field instability is likely to develop around the
magnetic field's symmetry axis (though this does not exclude the
possibility of it growing elsewhere within the star too). From a global
evolution, however, it is not 
straightforward to isolate the behaviour of the instability from
stable perturbations. Plots of $|\dB|$ across the numerical domain
can be dominated by the shape of an initially stable
perturbation, whereas we are interested in where
$|\dB|$ grows fastest \emph{after the onset of instability}. To monitor the
instability's growth, we divide the value of $|\dB|$ at each point 
by its value at some early time. We find that these plots then have a
generic structure similar to the one shown in figure \ref{B_pert}: the
perturbation grows fast around the axis, as expected, but also in a
shell towards the stellar surface. Comparing this with the shape of
the background field, shown in figure \ref{B_back}, we see that the
instability grows fastest where the background field is
weak. This resembles the situation for \emph{poloidal}-field
instabilities --- these grow fastest around the region
where the background field vanishes \citep{taylerpol}.

\begin{figure}
\begin{minipage}[c]{\linewidth}
\psfrag{growth}{$\zeta$/[s${}^{-1}$]}
\psfrag{Bav}{$\bar{B}$/[$10^{16}$ G]}
\psfrag{Om}{$\hat\Omega$}
\psfrag{B3}{$\Bav=3.0\times 10^{16}$ G}
\psfrag{B1.5}{$\Bav=1.5\times 10^{16}$ G}
\includegraphics[width=\linewidth]{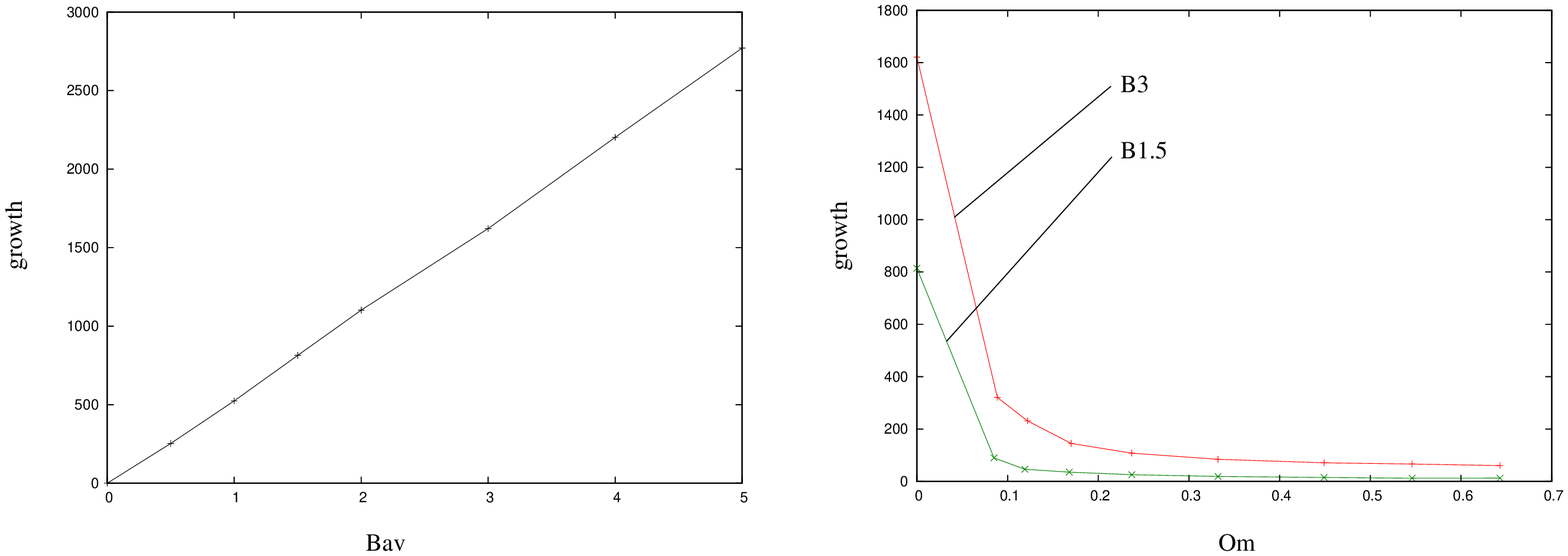}
\end{minipage}
\caption{\label{growthrates}
         Left: the instability growth rate (in seconds${}^{-1}$) plotted
         against field strength $\Bav$ for a series of nonrotating
         stars; we see that the dependence is linear for these field
         strengths. Right: The 
         effect of rotation on the growth rate, for two values of
         $\Bav$. At $\hat\Omega=0$ the weaker-field growth rate is
         half that of the stronger one; when $\hat\Omega>0$ it reduces
         to approximately a quarter that of the stronger one.}
\end{figure}

We now wish to quantify how the growth rate $\zeta$ of the Tayler
instability depends on field strength $\Bav$ and rotation rate
$\Omega$. For early times, the behaviour of the magnetic perturbations
will be a sum of various stable oscillatory modes and the unstable
perturbation $\propto \exp(kt)$ ($k\in\mathbb{R}_+$) which concerns
us; eventually, the amplitude of this unstable perturbation will become
dominant (see figure \ref{instab_rot}) and hence we may extract its
growth rate. We define 
\beq
\zeta=\frac{1}{\Delta t}\ \Delta\brac{\ln\brac{\frac{\delta M}{M_0}}}
\eeq
as a measure of growth, where $M_0$ is the magnetic energy of the
background. In figure \ref{growthrates} we use this to 
investigate how toroidal-field instabilities depend on the 
field strength and rotation rate of the star. In the left-hand plot we
show that $\zeta$ depends linearly on $\Bav$, whilst in the right-hand
plot we investigate the 
dependence of $\zeta$ on rotation rate $\Omega$, for $1.5$ and
$3.0\times 10^{16}$ G fields. At $\Omega=0$ $\zeta$ is half the value
for the weaker field (since it is linear in $\Bav$), but for
$\Omega>0$ the weaker field's growth rate $\zeta_{1.5}$ is about a
quarter that of the stronger one $\zeta_{3}$. This is consistent with
the suggestion of \citet{taylertor} that $\zeta$ should depend on the
ratio of magnetic to rotational energy $M/T$: if $\zeta_3\sim M_3/T$
and $\zeta_{1.5}\sim M_{1.5}/T$, then $\zeta_3/\zeta_{1.5}\sim
M_3/M_{1.5}\sim B_3^2/B_{1.5}^2=3^2/1.5^2 = 4$. We
note that rotation reduces the instability, but does not appear to
stabilise the field entirely; this is in agreement with the work of
\citet{pitts_tayler} and \citet{kitch}.

We conclude with a brief mention of $m\geq 2$ instabilities. In
\citet{tor_mode} we employed an artificial resistivity term to damp
out long-timescale instabilities in our evolutions. We returned to
study those instabilities for this work, but found that their growth
rates did not converge satisfactorily with resolution; hence, we can
not convincingly claim to have seen evidence of higher-$m$
toroidal-field instabilities. This is not surprising, though, as these
are weaker than the $m=1$ instability.

\vspace{-0.2cm}
\section{Tayler-instability effects in magnetars}

Magnetars in the SGR class are notable for their great variability in
luminosity. In addition to their quiescent emission of $\sim 10^{35}$
erg s${}^{-1}$, they undergo periods of short bursts at
luminosities of up to $\sim 10^{42}$ erg s${}^{-1}$. Rarely, they
suffer enormously energetic `giant flares', during which their
luminosity exceeds $\sim 10^{44}$ erg s${}^{-1}$ ($\sim 10^{47}$ erg
s${}^{-1}$, in the case of SGR 1806-20) \citep{mereghetti}. It is not
clear what provides the trigger mechanism for, in particular, these
giant flares; a number of internal and magnetospheric processes
have been suggested (see, for example, \citet{TD_magflare} or
\citet{gill_heyl} and references therein).

The Tayler instability is one possible candidate for an internal
SGR-flare trigger, as discussed briefly in
\citet{TD_magflare}; see also the related work on magnetic-field
reconfiguration by \citet{ioka}. Suppose
a magnetar field is dominantly toroidal 
--- such a configuration could come from, for example, differential
rotation winding up a mixed field at the star's birth. The small
poloidal component will help to suppress the toroidal-field
instability discussed in this paper, but nonetheless the field could
be on the verge of an unstable regime. One could imagine some
dissipative process weakening the poloidal component, or some
relatively minor field rearrangement occurring near the symmetry axis
of the field. The resulting field configuration could then be
susceptible to the Tayler instability, causing an
exponentially-growing perturbation in the magnetic energy which
manifests itself as a giant flare at the star's surface.

If this mechanism were the trigger for SGR flares, what could we learn
about magnetar fields? Let us return to the left-hand plot in figure
\ref{growthrates}, where we plot the instability growth rate $\zeta$
against averaged-field strength $\Bav$ for nonrotating stars (this is
appropriate for magnetars, whose rotational frequencies are close to
zero). Now consider a growth \emph{timescale}
$\tau_{grow}=1/\zeta$ (in seconds) for the instability. Our results
give an empirical relation
\beq
\tau_{grow}=1.9\brac{\frac{\Bav}{10^{16}\ \mathrm{G}}}^{-1}\ \mathrm{ ms}
\eeq
--- i.e, $\tau_{grow}=0.019$ seconds for a $\Bav=10^{15}$ G field and
$\tau_{grow}=1.9$ ms for a $10^{16}$ G field. This latter value is
comparable with upper limits on the initial rise times for the three
observed giant flares: 1.3 ms (SGR 1806-20), 1.6 ms (SGR 1900+14) and
2 ms (SGR 0526-66) \citep{tanaka}. If we associate these rise times with
$\tau_{grow}$ then we obtain values for the averaged-field strength of
$0.95, 1.2$ and $1.5\times 10^{16}$ G for SGRs 0526-66, 1900+14 and
1806-20 respectively. These values seem reasonable in the context of
recent estimates \citep{stella_highB,kaminker_highB} for magnetar
internal field strengths.

Finally, if the trigger mechanism for giant flares resembles a large
$m=1$ perturbation of the magnetic field, as we suggest here, it would
induce a corresponding disturbance in the stellar density distribution. The
$l\geq 2$ harmonics in this disturbance will produce gravitational
radiation --- making SGR giant flares a suitable, if rare, target for
gravitational-wave burst searches from advanced LIGO. We are, however,
unable to estimate the amplitude of these bursts, since nonlinear
effects (not included in our work) would become important after the
initial rise time.

\vspace{-0.2cm}
\section{Discussion}

In this paper we have shown that the unstable nature of purely
toroidal magnetic fields, well-known from local analyses about the
magnetic axis, may also be seen from a global evolution of linear
perturbations. In particular, a generic configuration that emerges
when studying axisymmetric Newtonian MHD --- where $B_\phi$ is
proportional to $\rho r\sin\theta$ --- proves to be unstable to $m=1$
perturbations. The instability is reduced, but not eliminated, by
rotation; this agrees with the suggestion of \citet{pitts_tayler} and
the studies by \citet{kiuchi_evol} and
\citet{kitch}. \citet{braithtor} found rotation could entirely
stabilise certain toroidal-field configurations. We have quantified
the dependence of the instability growth rate on field strength and
rotation rate.

The Tayler instability rules out generic purely-toroidal
configurations as candidates for physical stellar fields in strict
equilibrium (at least in barotropic-fluid stars; see
\citet{reis_strat} for arguments that it may be a poor assumption to
regard stars as barotropic). The instability could however be of
astrophysical interest: if a stellar magnetic field is dominantly
toroidal it is likely it will be close to unstable, with relatively
minor disturbances able to trigger the instability. We have discussed
the possibility of such a mechanism causing the giant flares of SGRs,
and motivation for gravitational-wave burst searches coincident with
these giant flares.

Although the Tayler instability may be relevant in explaining SGR
flares, we are many steps away from a credible study of magnetic-field
stability in neutron stars. Natural topics for further studies would be
the stability of purely poloidal fields and mixed-field
configurations, which we intend to consider in future work. Further
into the future, one might hope to consider the effect of
superconductivity and/or an elastic crust, and so build up a
more realistic picture of neutron star stability.

\vspace{-0.3cm}
\section*{Acknowledgments}

SKL acknowledges funding from a Mathematics Research Fellowship from
Southampton University. This work was supported by STFC through grant
number PP/E001025/1 and by CompStar, a Research Networking Programme
of the European Science Foundation. We thank Kostas Glampedakis for
helpful comments on a draft of this paper.

\label{lastpage}


\vspace{-0.3cm}


\begin{thebibliography}{99}
\bibitem[\protect\citeauthoryear{Andersson \& Kokkotas}{2001}]{nils_rmode}
  Andersson N., Kokkotas K.D., 2001, International Journal of Modern
  Physics D, 10, 381
\bibitem[\protect\citeauthoryear{Arfken \& Weber}{2001}]{arfken}
  Arfken G.B., Weber H.J., 2001, Mathematical Methods for Physicists,
  Harcourt Academic Press
\bibitem[\protect\citeauthoryear{Braithwaite}{2006}]{braithtor}
  Braithwaite J., 2006, A\&A, 453, 687
\bibitem[\protect\citeauthoryear{Christensen-Dalsgaard \&
    Gough}{2001}]{dipfmode}
  Christensen-Dalsgaard J., Gough D.O., 2001, MNRAS, 326, 1115
\bibitem[\protect\citeauthoryear{Dedner et al.}{2002}]{dedner}
  Dedner A., Kemm F., Kr\"oner D., Munz C.-D., Schnitzer T., Wesenberg
  M., 2002, J. Comp. Phys., 175, 645
\bibitem[\protect\citeauthoryear{Dib, Kaspi \& Gavriil}{2007}]{dib}
  Dib R., Kaspi V.M., Gavriil F.P., 2007, ApJ, 673, 1044
\bibitem[\protect\citeauthoryear{Duncan \& Thompson}{1992}]{duncthom}
  Duncan R.C., Thompson C., 1992, ApJ, 392, L9
\bibitem[\protect\citeauthoryear{Gill \& Heyl}{2010}]{gill_heyl}
  Gill R., Heyl J.S., 2010, arXiv:1002.3662
\bibitem[\protect\citeauthoryear{Glampedakis \& Andersson}{2007}]{glam_ander}
  Glampedakis K., Andersson N., 2007, MNRAS, 377, 630
\bibitem[\protect\citeauthoryear{Goossens \& Tayler}{1980}]{goostayler}
  Goossens M., Tayler R.J., 1980, MNRAS, 193, 833
\bibitem[\protect\citeauthoryear{Ioka}{2001}]{ioka}
  Ioka K., 2001, MNRAS, 327, 639
\bibitem[\protect\citeauthoryear{Kaminker et al.}{2007}]{kaminker_highB}
  Kaminker A.D., Yakovlev D.G., Potekhin A.Y., Shibazaki N.,
  Shternin P.S., Gnedin O.Y., 2007, Ap\&SS, 308, 423
\bibitem[\protect\citeauthoryear{Kitchatinov \& R\"udiger}{2008}]{kitch}
  Kitchatinov L.L., R\"udiger G., 2008, A\&A, 478, 1
\bibitem[\protect\citeauthoryear{Kiuchi, Shibata \& Yoshida}{2008}]{kiuchi_evol}
  Kiuchi K., Shibata M., Yoshida S., 2008, PRD, 78, 024029
\bibitem[\protect\citeauthoryear{Lander \& Jones}{2009}]{landerjones}
  Lander S.K., Jones D.I., 2009, MNRAS, 395, 2162
\bibitem[\protect\citeauthoryear{Lander, Jones \& Passamonti}{2010}]{tor_mode}
  Lander S.K., Jones D.I., Passamonti A., 2010, MNRAS, 405, 318
\bibitem[\protect\citeauthoryear{Lockitch \& Friedman}{1999}]{lock_fried}
  Lockitch K.H., Friedman J.L., 1999, ApJ, 521, 764
\bibitem[\protect\citeauthoryear{Markey \& Tayler}{1973}]{taylerpol}
  Markey P., Tayler R.J., 1973, MNRAS, 163, 77
\bibitem[\protect\citeauthoryear{Mereghetti}{2008}]{mereghetti}
  Mereghetti S., 2008, A\&A Rev., 15, 225
\bibitem[\protect\citeauthoryear{Miralles, Pons \& Urpin}{2002}]{miralles}
  Miralles J.A., Pons J.A., Urpin V.A., 2002, ApJ, 574, 356
\bibitem[\protect\citeauthoryear{Pitts \& Tayler}{1985}]{pitts_tayler}
  Pitts E., Tayler R.J., 1985, MNRAS, 216, 139
\bibitem[\protect\citeauthoryear{Price \& Monaghan}{2005}]{pricemona}
  Price D.J., Monaghan J.J., 2005, MNRAS, 364, 384
\bibitem[\protect\citeauthoryear{Reisenegger}{2009}]{reis_strat}
  Reisenegger A., 2009, A\&A, 499, 557
\bibitem[\protect\citeauthoryear{Rezzolla, Lamb \& Shapiro}{2000}]{rezz_rmode}
  Rezzolla L., Lamb F.K., Shapiro S.L., 2000, ApJ, 531, L139
\bibitem[\protect\citeauthoryear{Roxburgh}{1963}]{rox_tor}
  Roxburgh I.W., 1963, MNRAS, 126, 67
\bibitem[\protect\citeauthoryear{Roxburgh \& Durney}{1967}]{rox_durn}
  Roxburgh I.W., Durney B.R., 1967, MNRAS, 135, 329
\bibitem[\protect\citeauthoryear{Stella et al.}{2005}]{stella_highB}
  Stella L., Dall'Osso S., Israel G.L., Vecchio A., 2005, ApJ, 634, L165
\bibitem[\protect\citeauthoryear{Tanaka et al.}{2007}]{tanaka}
  Tanaka Y.T., Terasawa T., Kawai N., Yoshida A., Yoshikawa I., Saito
  Y., Takashima T., Mukai T., 2007, ApJ, 665, L55
\bibitem[\protect\citeauthoryear{Tayler}{1973}]{taylertor}
  Tayler R.J., 1973, MNRAS, 161, 365
\bibitem[\protect\citeauthoryear{Thompson \& Duncan}{1996}]{TD_magflare}
  Thompson C., Duncan R.C., 1996, ApJ, 473, 322
\bibitem[\protect\citeauthoryear{Yoshida \& Lee}{2000}]{yoshlee}
  Yoshida S., Lee U., 2000, ApJ, 529, 997
\end{thebibliography}
\end{document}